\documentclass{iaus}
\usepackage{graphicx}

\title[Galaxy Environments at $z \sim 1$] %% give here short title %%
{\large Galaxy Environments in DEEP2: The Birth of the Red Sequence}

\author[Cooper \& Newman]   %% give here short author list %%
{Michael C.\ Cooper, Jeffrey A.\ Newman}

\affiliation{Dep't of Astronomy, University of California at Berkeley,
Berkeley, CA 94720, USA 
\break email: cooper@astro.berkeley.edu; jnewman@astro.berkeley.edu}

\pubyear{2006}
\volume{IAUS235}  %% insert here IAU Symposium No.
\pagerange{100--101}
\date{?? and in revised form ??}
\jname{Proceedings Title IAU Symposium}
\editors{A.C. Editor, B.D. Editor \& C.E. Editor, eds.}
\begin{document}

\maketitle

\firstsection % if your document starts with a section,
\section{Introduction}

The galaxy population at $z \lesssim 1$ is effectively described as a
combination of two distinct types: red, early-type galaxies lacking much
star formation and blue, late-type galaxies with active star formation. For
the red galaxy population, recent work by \cite{bell04} has shown that the
number density of $\sim \! L^{*}$ galaxies on the red sequence has risen by
a factor of $\sim \! 2$ from $z \sim 1$ to $z \sim 0$. A variety of
complementary observations suggests that the build-up of galaxies on the
red sequence results from 2 distinct evolutionary trends: (1) the quenching
of star formation in blue galaxies and their subsequent migration onto the
red sequence and (2) the dissipationless or (``dry'') merging of
red-sequence galaxies.

Many of the physical processes responsible for the quenching of star
formation have a strength which depends on environment. For example,
ram-pressure stripping and galaxy harassment operate most effectively in
regions of extremely high galaxy overdensity (i.e., galaxy clusters). Thus,
within this picture, the formation of red-sequence galaxies is directly
connected to the local environment. Further evidence for a strong
connection between environment and the formation of early-type galaxies is
apparent in what are termed the color-density and mophology-density
relations. As first quantified more than 25 years ago, these relations hold
that star-forming, disk-dominated galaxies tend to reside in regions of
lower galaxy density relative to those of red, bulge-dominated galaxies.

Ultimately, we aim to understand when the build-up of red-sequence galaxies
began and what specific processes drove this evolution in the galaxy
population. The latter issue is one discussed by many presenters in this
IAU symposium (e.g., by Conselice, Zhang, etc.). Here, we focus our
attention on determining both \emph{where} and \emph{when} the red sequence
was first established. Did the typical early-type galaxy cease star
formation at very early times ($z \sim 3$), or did they quench recently (in
the last 7 Gyr)?

\section{DEEP2 and the Evolution of  the Color-Density Relation}\label{sec:deep2}

The DEEP2 Galaxy Redshift Survey began spectroscopic observations in the
Summer of 2002, with the goal of characterizing the galaxy population and
large-scale structure at $z \sim 1$. To date, the survey has targeted $\sim
\! 50,000$ galaxies in the redshift range $0.2 < z < 1.4$, down to a
limiting magnitude of $R_{\rm AB} = 24.1$. The second major data release
for DEEP2 is scheduled for early 2007, including approximately 75\% of the
data.

Herein, we utilize a sample of $> \! 10,000$ galaxies with redshifts in the
range $0.75 < z < 1.35$ and drawn from all 4 of the DEEP2 survey fields. To
enable comparison over a broad $z$ range, we restrict analyses to include
equivalent populations of galaxies at all $z < 1.3$. Details of this
selection are discussed by \cite{cooper06b} and \cite{gerke06b}. For each
galaxy in the DEEP2 sample, we quantify the environment in terms of the
local galaxy overdensity $(1 + \delta_3)$, which is derived from the
projected distance to the $3^{\rm rd}$-nearest-neighbor, as detailed by
\cite{cooper06a}. In the tests of \cite{cooper05}, this environment
estimator proved to be the most robust indicator of local galaxy density
for intermediate-$z$ surveys like DEEP2.

Within the DEEP2 sample, a bimodal distribution of galaxies in rest-frame
$U-B$ color is clearly visible out to $z > 1$, with galaxies populating a
relatively tight red sequence and a more diffuse blue cloud. Here, we
compute the fraction of galaxies on the red sequence using the color
divison defined by Eq.\ 19 of \cite{willmer06}, where the red fraction
$(f_{\rm R})$ within a given redshift and environment range is given by the
number of galaxies redward of this relation in $U-B$ color divided by the
total number of galaxies within the same bin of redshift and environment.

\begin{figure}
\parbox{2.8in}{
\includegraphics[angle=0,width=2.8in]{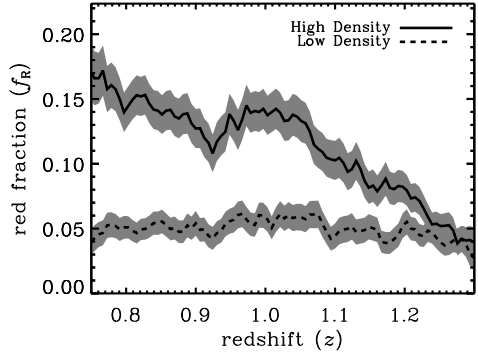}
}
\parbox{2.0in}{ \vskip -0.2in 

  \caption{For galaxies within high-density (\emph{solid line}) and
  low-density (\emph{dashed line}) environments, we plot the red fraction
  $(f_{\rm R})$ as a function of redshift for galaxies in sliding bins of
  $\Delta z = 0.1$. The grey shaded regions give the 1--$\sigma$ range of
  the red fractions in each density regime.} }
\label{fig:color-density}
\end{figure}

To study the evolution of the color-density relation with $z$ in detail, we
divide the DEEP2 sample into thirds according to overdensity and compute
$f_{R}$ as a function of redshift for the galaxies in the high-density and
low-density extremes of the overdensity distribution, which minimizes
cross-contamination. As illustrated in Fig.\ \ref{fig:color-density}, the
$f_{\rm R}$-overdensity relation shows a continuous evolution at $0.75 < z
< 1.3$, such that at $z \sim 1.3$ the red fraction in low- and high-density
regions are statistically consistent with each other. Extrapolating linear
regression fits to the $f_{\rm R}(z)$ relations in high-density and
low-density environments, we find convergence at $z = 1.33$. Tests
utilizing mock galaxy catalogs show that this observed evolution is not due
to any observational selection effect. Additionally, the results presented
in Fig.\ \ref{fig:color-density} are unchanged when using highly disimilar
sample selections and only minimally impacted by uncertainties due to
cosmic variance.

The data presented in this proceeding consititute the first robust
measurements of the color-density relation at $z > 1$, spanning
environments from rich groups down to voids. We find that the clustering of
$\sim \! L^{*}$ galaxies at $z \gtrsim 1.3$ should depend only weakly on
color. Furthermore, we conclude that the build-up of the red sequence is
largely driven by galaxy environment, with the first $L^{*}$ galaxies
quenching and moving on to the red sequence at $z \sim 1.7$. This epoch
marks the birth of the red sequence, the point at which typical $L^{*}$
galaxies began quenching in significant number, while at earlier times the
red galaxy population is comprised of only the extremely massive and rarest
systems.

\begin{acknowledgments}
  This work was supported in part by NSF grant
  AST0507428.
\end{acknowledgments}

\end{document}